\documentclass[preprint]{revtex4}
\usepackage{amsmath}
\usepackage{graphicx}
\usepackage{psfrag}
\usepackage{dcolumn}
\usepackage{bm}
\usepackage{bm}
\usepackage{color}
\usepackage{wasysym}
\usepackage{amssymb}

\newcommand{\DEF}{\stackrel{\mbox{\rm\scriptsize def}}{=}}
\renewcommand{\d}{\mathrm{d}}

\begin{document}

\title{Mechanical response of plectonemic DNA:\\
an analytical solution}

\author{N. Clauvelin}
\author{B. Audoly}
\author{S. Neukirch}
\affiliation{%
UPMC Univ Paris 06 \& CNRS\\
UMR 7190, Institut Jean le Rond d'Alembert,\\
Paris, France}
\date{\today}
\begin{abstract}
We consider an elastic rod model for twisted DNA in the plectonemic
regime.  The molecule is treated as an impenetrable tube with an
effective, adjustable radius.  The model is solved analytically and we
derive formulas for the contact pressure, twisting moment and
geometrical parameters of the supercoiled region.  We apply our model
to magnetic tweezer experiments of a DNA molecule subjected to a
tensile force and a torque, and extract mechanical and geometrical
quantities from the linear part of the experimental response curve.
These reconstructed values are derived in a self-contained manner, and
are found to be consistent with those available in the literature.
\end{abstract}


%
\maketitle








%
%
%
\section{Introduction}
%
%
%
%
%
%
%
%

Mechanical properties of the DNA molecule play an important role in
the biological processes involved in the cell, yet we only have an
imprecise view of these properties.  Advances in nanotechnologies
make it possible to exert forces onto isolated DNA filaments:
mechanical response of the molecule is now widely studied.
Single molecule experiments provide a powerful way to investigate the
behavior of DNA subjected to mechanical stress.  In such experiments,
the molecule is held by optical or magnetic tweezers and forces and
torques are applied to it~\cite{smith+al:1992,bustamente+al:1994}.
The interaction between DNA and proteins is actively investigated; for
instance, the chemical and mechanical action of an enzyme on a
molecule can be inferred from the global deformation of the
molecule~\cite{RevyakinLiuEbright-Abortive-Initiation-and-Productive-Initiation-by-RNA-Polymerase-2006}.

In this paper we focus on a specific type of experiments: a double
stranded DNA molecule is fixed by one end to a glass surface while the
other end is attached to a magnetic bead; using a magnet, a pulling
force and a torque are applied on the DNA
filament~\cite{strick+al:1996}.
Large ranges of pulling forces, from one tenth to one hundred
piconewton, and number of turns can be explored in the experiments,
and the molecule displays a variety of behaviors and
conformations~\cite{allemand+al:1998,sarkar+al:2001,bryant+al:2003}.
We study the response of the molecule to moderate forces, below 10~pN,
and moderate to large number of turns, equivalent to a positive
supercoiling ratio of the order of 0.1.
In experiments, the pulling force is kept constant while the bead is
rotated gradually.  Above a threshold value of the
number of turns, the molecule wraps around itself in a helical way,
giving rise to a structure comprising plectonemes.
The vertical extension of the molecule is recorded and plotted as a
function of the number of turns.
Experimental rotation-extension curves have a characteristic shape and
are called \emph{hat curves}~\citep{bouchiat_2000, strick_1998}.  The
central, rounded part of the curve can be explained using the
worm-like chain (WLC) model~\citep{marko+siggia:1995a} and its
variants.  At larger number of turns, the extension of the molecule
decreases linearly.
This linear part is obtained when the molecule is in supercoiled
configuration and forms plectonemes.
The plectonemic structure is made of two interwound helical filaments
whose geometry is characterized by the so-called superhelical angle
and radius; note that each of these filaments is itself made of a
double-stranded DNA molecule.  The superhelical angle and the
twisting moment in the filaments are key parameters that control the
action of topoisomerases~\citep{koster+al:2005}, RNA
polymerase~\cite{revyakin+al:2004}, or other
enzymes~\cite{heijden+al:2005} on DNA.
The distance of self-approach of DNA in supercoiled regime has been
the subject of a number of
studies~\cite{bednar+al:1994,rybenkov_1997,rybenkov+al:1993,charvin+al:2005}.
In previous analytical and numerical work, the double stranded DNA
molecule has been modelled as a twist-storing elastic filament.  These
approaches have been successful at reproducing the response of DNA to
moderate torque~\citep{bouchiat_1998,moroz+nelson:1997}, given by the
central region of the experimental curves.  The analysis of the linear
regions of these curves, based on a detailed model of plectonemes, was
lacking until recently: in Ref.~\cite{marko:2007}, a composite model
based on an empirical free energy of supercoiled DNA is proposed.

Here we present an elastic rod model for helical supercoiling of the
DNA molecule, which is relevant to a large number of turns.
Our model is self-contained and provides an mechanically accurate
description of elastic filaments in contact.  The molecule is divided
in two domains: one where the configuration is a worm-like-chain,
dominated by thermal fluctuations, and the other one, a superhelical
region dominated by elasticity, where the molecule contacts itself.
The plectonemic regions can be spread in various places of the
molecule; as this does not change the mechanical response of the
system, we refer to these regions as if they were in one chunk.
We deal with self-contact by introducing an effective superhelical
radius (distinct from the crystallographic radius of 1~$\textrm{nm}$,
from the size of the Manning condensate and from the Debye length,
although in the same range of values), which varies with external
loads and salinity of the solution.  The effective radius is defined
as the radius of a chargeless, impenetrable and elastic tube having
the same mechanical response as the molecule.  This radius is not
given as a parameter of the model and is extracted from experimental
data.
Using an energy approach, we relate geometrical variables
(superhelical radius and angle) to applied force and torque.  We also
characterize the response of the molecule in the plectonemic regime,
extend former numerical results~\cite{neukirch_2004}, and show how
geometrical and mechanical parameters can be extracted from
experimental data.

%
%
%
\section{Model}
%
%
%
%
%
%
%
%

The present model investigates the equilibrium behavior of an elastic
rod with bending rigidity $K_0$ (the bending persistence length is
$A=K_0/(kT)$, where $k$ is the Boltzmann constant and $T$ the absolute
temperature) and twisting rigidity $K_3$ under traction and torsion as
shown in Fig.~\ref{fig:ExperimentSetup}.
This is a coarse-grained model for DNA where base-pairs details are
neglected.  For instance, the anisotropic flexibility of the molecule,
originating from base pairing and major-minor grove geometry, is
smoothed out at a scale of several base pairs: a highly twisted
anisotropic rod can be replaced by an equivalent isotropic rod with
effective bending rigidity~\citep{kehrbaum_2000}.

\subsection*{Geometry}
\begin{figure}
    \begin{center}
 	    \includegraphics[width=9.cm]{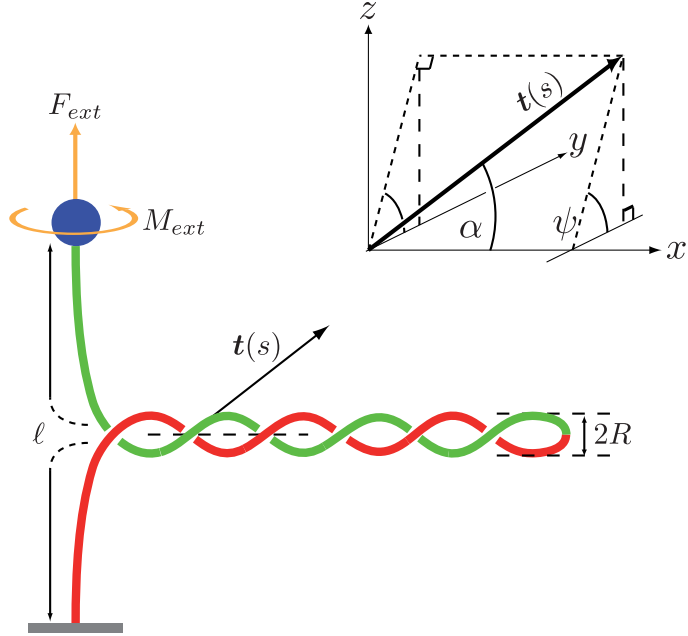}
	\caption{Sketch of the magnetic tweezers experiment.  A B-DNA
	molecule of total contour length $\ell$ is fixed in $s=0$ to a
	glass surface while the other end in $s=\ell$ is attached to a
	magnetic bead.  A pulling force $F_{\mathrm{ext}}$ and a
	torque $M_{\mathrm{ext}}$ are applied at the upper end by
	using a magnet.  The superhelical angle and radius are denoted
	$\alpha$ and $R$ respectively.}
	\label{fig:ExperimentSetup}
    \end{center}
\end{figure}

%
%
%
%
%
We start with a geometric description of the rod configurations
relevant to the plectonemic regime.  This defines a reduced set of
configurations (Ansatz), over which we shall minimize the
elastic strain energy associated with deformations.
The rod, of length $\ell$, is considered inextensible and has circular
cross-section; let $s$ denote the arclength along the rod.  The strain
energy involves, at lowest order, the geometric curvature $\kappa(s)$
of the centerline of the rod as well as the twist $\tau(s)$.
We emphasize that the twist $\tau(s)$ is different from the
geometrical (Fr\'enet-Serret) torsion of the centerline as it takes
into account the rotation of material cross sections around the
centerline.  It allows one to distinguish between twisted and
untwisted configurations of the rod having the same centerline.
The rod centerline is parameterized by $\mathbf{r}(s)$ and its unit
tangent $\mathbf{t} \DEF \d \mathbf{r} / \d s $ can be described with
spherical angles, as shown in Fig.~\ref{fig:ExperimentSetup}:
$\alpha(s)$ is the zenith angle and $\psi(s)$ the azimuth angle with
respect to the direction $\mathbf{e_x}$ along the common axis of the two
superhelices in the plectonemic region.

We consider the following configurations, relevant to a large applied
number of turns, $n$.
The tails are assumed to be straight but twisted (thermal fluctuations
will be accounted for by using the rescaled tail length predicted by
WLC theory).
The plectonemes are described by two identical and uniform helices
where, again, each of these helices is itself a double-stranded DNA
molecule.
Both the end loop of the plectonemes and the matching region between
the tails and the plectonemic part are neglected.

In the tails the rod is straight and aligned with the $\mathbf{e_z}$
axis, and $\mathbf{t}=\mathbf{e_{z}}$ there.  The geometric curvature
$\kappa \DEF \left| \d \mathbf{t} / \d s \right|$ is zero,
$\kappa(s)=0$.  

In each filament of the plectonemes, the position vector $\mathbf{r}(s)$
and the tangent vector $\mathbf{t}(s)$ describe a superhelix of axis
$\mathbf{e_x}$:
\begin{equation}
	\begin{cases}
		r_x(s)=s\,\cos\alpha \\
		r_y(s)=\chi \,R\,\sin\psi(s) \\
		r_z(s)=-\chi \,R\,\cos\psi(s)
	\end{cases}
	\; \text{and} \;\;
	\begin{cases}
		t_x(s)=\cos\alpha \\
		t_y(s)=\sin\alpha\,\cos\psi(s) \\
		t_z(s)=\sin\alpha\,\sin\psi(s) 
	\end{cases}
	\label{eqn:PosVector-Plecto}
\end{equation}
The other filament of the plectonemes is obtained by a rotation of
$180^\circ$ around the axis $\mathbf{e_{x}}$.  Here $\chi=\pm 1$
stands for the chirality of the two helices and $R$ and $\alpha$
denote the superhelical radius and angle, respectively.
In equation~(\ref{eqn:PosVector-Plecto}), the condition
$\mathrm{d}\mathbf{r}/\mathrm{d}s = \mathbf{t}$ yields $ \d \psi / \d
s=\chi \, \sin\alpha / R $.
The curvature in the plectonemes is $\kappa(s) \DEF \left| \d
\mathbf{t} / \d s \right| = \frac{\sin^2\alpha}{R}$.

Noting $\ell_{\mathrm{p}}$ the contour length spent in the plectonemes,
we obtain the following expression for the integral of the squared curvature 
over the whole length of the rod:
\begin{equation}
    \int_{0}^{\ell} \kappa^2\,\mathrm{d}s = 
    \frac{\sin^4\alpha}{R^2}\,
    \ell_{\mathrm{p}}
    \textrm{.}
    \label{eq:IntegralOfSquareCurvature}
\end{equation}

The end torque twists the filament.  For a rod with circular
cross-section, the twist $\tau(s)$ at equilibrium is
uniform~\citep{antman_2005}, $\d \tau/ \d s = 0$ for all $s$.  As a
result, the equilibrium configuration of the rod is fully specified by
the centerline, through the variables $\alpha$, $R$ and
$\ell_{\mathrm{p}}$, and an additional scalar $\tau$ describing twist.

The twist parameter $\tau$ is geometrically related to the number of
turns imposed on the magnetic bead, $n$, which is equal to the link of
the DNA molecule, $n=\mathrm{Lk}$.  In the present case the link
reads~\citep{neukirch_2004}:
\begin{equation}
	\mathrm{Lk} = \mathrm{Tw}+\mathrm{Wr} =
	\frac{1}{2\pi}\int^\ell_0\tau\,\d s -
	\chi\,\frac{\sin2\alpha}{4\pi\,R}\,\ell_p
	=\frac{1}{2\pi}\,\left(
	\tau\,\ell
	-\chi\,\frac{\sin2\alpha}{2\,R}\,\ell_{\mathrm{p}}
	\right),
	\label{eqn:LinkDefinition}
\end{equation}
as we neglect the writhe in the tails.

\subsection*{Energy formulation}
%
%
%
%
%
%
Using the above notations the rod is described by four variables:
$\alpha$ the superhelical angle, $R$ the superhelical radius, $\tau$
the twist and $\ell_p$ the contour length spent in the plectonemes.
We proceed to derive the total energy of the system as a function of
these four variables.  It is the sum of three terms,
$V=V_{\mathrm{el}}+V_{\mathrm{ext}} + V_{\mathrm{int}}$, where the first term is the
strain elastic energy, the second is the potential energy
associated with the external loads $F_{\mathrm{ext}}$ and
$M_{\mathrm{ext}}$, and the third accounts for interaction of the
filaments in the plectonemes.
The strain elastic energy for the rod of total contour length $\ell$
is :
\begin{equation}
	V_{\mathrm{el}}=\frac{K_0}{2}\int^{\ell}_{0}\kappa^2\,\d
	s+\frac{K_3}{2}\int^{\ell}_{0}\tau^2\,\d s~.
	\label{eqn:ElasticEnergy}
\end{equation}
We do not take into account the reduction of the effective torsional
rigidity in the tails due to fluctuations~\cite{moroz+nelson:1997}.
The potential energy is given by:
\begin{equation}
	V_{\mathrm{ext}}=-F_{\mathrm{ext}} (z(\ell)-z(0)) -
	2\pi\,M_{\mathrm{ext}}\,n~,
	\label{eqn:LoadsEnergy}
\end{equation}
where $z(\ell)-z(0)=\ell - \ell_p$ for straight tails and
$n=\mathrm{Lk}$.

If the DNA-DNA interaction was clearly established, we
would include the corresponding interaction energy $V_{\mathrm{int}}$
in the total energy $V$~\cite{CAN_U_electro}.
This is not the case and we model the filaments in electrostatic
interaction as effective chargeless hard-core tubes.  The effective
radius $a$ of these tubes accounts for a variety of physical
mechanisms and in particular for the presence of counter-ions.
As in Refs.~\cite{VologodskiiCozzarelli-Modeling-of-long-range-electrostatic-interactions-in-DNA-1995,%
rybenkov_1997}, we do not fix the quantity $a$ in
advance and let it vary with experimental conditions, such as applied
load and salinity.  In fact, we show how $a$ can be extracted from
experimental measurements.  Doing so, we replace the actual (unknown)
interaction potential $V_{\mathrm{int}}(R,\alpha)$ by a hard-core
interaction with adjustable radius $a$, and optimize $a$ to best fit a
given experiment.  

The parameter $a$ must certainly be larger than the crystallographic
DNA radius $1~\mathrm{nm}$.  It is different from the radius of the
Manning condensate~\cite{manning_1969_1,manning_1969_2,manning_1969_3}
since approximately a quarter of the charge remains outside of the
Manning condensate.
The equilibrium is the solution of a constrained minimization problem
for the elastic energy, subjected to the impenetrability condition
\begin{equation}
    R\geq a
    \textrm{.}
    \label{eq:RLargerThanA}
\end{equation}
We anticipate on the fact that there is contact, $R=a$, for typical
experimental conditions.  This constraint is handled by a Lagrange
multiplier $\lambda$; the actual interaction energy is then
substituted with the following expression:
\begin{equation}
	V_{\mathrm{int}}=-\lambda\left(R-a\right)
	\textrm{.}
	\label{eqn:InteractionEnergy}
\end{equation}

Combining
Eqs.~(\ref{eq:IntegralOfSquareCurvature}--\ref{eqn:InteractionEnergy}),
we write the total potential energy of the system as:
\begin{multline}
	V(\alpha, R, \tau, \ell_{\mathrm{p}}) =
	\frac{K_0}{2}\,\frac{\sin^4\alpha}{R^2}\,\ell_p +
	\frac{K_3}{2}\,\tau^2\,\ell
	- F_{\mathrm{ext}}\left(\ell-\ell_\mathrm{p}\right) 
	\\ {}-M_{\mathrm{ext}}\,\left(
	\tau\,\ell
	-\chi\,\frac{\sin2\alpha}{2\,R}\,\ell_{\mathrm{p}}
	\right)
	- \lambda\left(R-a\right)
	\textrm{.}
	\label{eqn:TotalEnergy}
    \end{multline}
In Ref.~\cite{kutter_2001} a similar energy function has been
introduced but the rest of analysis differs from ours.  Indeed, their
approach focuses on statistical mechanics and the analysis of the
state of lowest energy is overlooked.  Moreover, the parameter $a$ is
fixed a priori to the crystallographic radius of DNA,
$a=1~\mathrm{nm}$, which is a strong underestimation of the actual
distance of self-approach of DNA in saline solution.  In contrast, we
undertake a detailed analysis of the equilibrium solutions, with
thermal fluctuations considered in the tails; this allows us to derive
simple formulas for the force and the moment as a function of the
superhelical variables, applicable to magnetic tweezers experiments.

%
%
\section{Results}
%
%
%
%
%
%
Mechanical equilibrium is given by the Euler-Lagrange condition for
the stationarity of the potential $V(\alpha,
R,\tau,\ell_{\mathrm{p}})$ in Eq.~(\ref{eqn:TotalEnergy}) with respect
to its variables,
\begin{equation*}
	\left( 
	\frac{\partial V}{\partial \tau},\, 
	\frac{\partial V}{\partial \alpha},\,
	\frac{\partial V}{\partial \ell_\mathrm{p}},\, 
	\frac{\partial V}{\partial R} \right)=
	0.
\end{equation*}

The first condition $\partial V/\partial \tau$ allows one to recover
the constitutive relation for twist deformations, $M_{\mathrm{ext}} =
K_{3}\, \tau$, given that the twisting moment is uniform in the
filament and equal to the applied torque $M_{\mathrm{ext}}$.

Variation of the total energy with respect to $\alpha$ gives the
expression of the applied torque $M_{\mathrm{ext}}$ in terms of the
superhelical variables $\alpha$ and $R$:
\begin{equation}
	M_{\mathrm{ext}} =
	-\frac{2\chi\,K_0}{R}\frac{\cos\alpha\,\sin^3\alpha}{\cos2\alpha}
	\textrm{,}
	\label{eqn:VarTorsionalMoment}
\end{equation}
which is what was found for purely plectonemic solution (no
tails)~\cite{thompson+al:2002}.

The condition $\partial V/\partial \ell_{\mathrm{p}} = 0$, combined
with Eq.~(\ref{eqn:VarTorsionalMoment}), allows one to relate the
pulling force $F_{\mathrm{ext}}$ to the superhelical geometry:
\begin{equation}
	F_{\mathrm{ext}} =
	\frac{K_0}{R^2}\,\sin^4\alpha\,\left(\frac{1}{2}+\frac{1}{\cos 2\alpha}\right)
	\textrm{.}
	\label{eqn:VarPullingForce}
\end{equation}
This formula justifies and extends the numerical fit $F_{\mathrm{ext}}
\propto K_{0}\,\alpha^4/R^2$ found in Ref.~\cite{neukirch_2004} for
small values of $\alpha$.

The Euler-Lagrange condition with respect to $R$ yields an equation
involving the Lagrange multiplier $\lambda$.  The quantity $\lambda /
\ell_{\mathrm{p}}$ can be interpreted as the contact force per unit
length, $p$, of one filament onto the other.
Eqs.~(\ref{eqn:TotalEnergy}--\ref{eqn:VarTorsionalMoment}), together
with the condition $\partial V/\partial R=0$, yields:
\begin{equation}
p=\frac{\lambda}{\ell_p}=\frac{K_0}{R^3}\frac{\sin^4\alpha}{\cos2\alpha}
\textrm{.}
	\label{eqn:VarContactPressure}
\end{equation}
Note that this pressure (more accurately, force per unit length) is
positive for $\alpha\leq \pi/4$; if our assumption of contact $R=a$
was incorrect, this would be indicated by a negative pressure value here.

In magnetic tweezers experiments, the pulling force $F_{\mathrm{ext}}$
is imposed although the applied torque $M_{\mathrm{ext}}$ is unknown.
The two unknowns $R$ and $\alpha$ are then related by
Eq.~(\ref{eqn:VarPullingForce}); in the next Section, a second
equation relating those unknowns and the extension $z$ is given, which
makes it possible to solve for $R$ and $\alpha$.  The twisting moment
can then be found from Eq.~(\ref{eqn:VarTorsionalMoment}).

\subsection*{Vertical extension of the filament}
%
%
%
%
In magnetic tweezers experiments, the measurable quantities are the
vertical extension $z$ and the number of turns $n$ imposed on the
bead.
Using Eq.~(\ref{eqn:LinkDefinition}) for $n=\mathrm{Lk}$, the equation
$z = \ell-\ell_\mathrm{p}$ and the constitutive relation $\tau =
M_{\mathrm{ext}}/K_{3}$ where $M_{\mathrm{ext}}$ is found from
Eq.~(\ref{eqn:VarTorsionalMoment}), we obtain the vertical extension
of the filament as a linear function of the number of turns $n$:
\begin{equation}
	z =
	\left(1+\frac{2\,K_0}{K_3}\frac{\sin^2\alpha}{\cos2\alpha}\right)\ell
	+\chi\,n\,\frac{4\pi\,R}{\sin2\alpha}~.
	\label{eqn:VerticalExtension}
\end{equation}
Thermal fluctuations dominantly affect the tails and make the
end-to-end distance $z$ of the molecule smaller than the contour
length $\ell-\ell_{\mathrm{p}}$ of the tail parts, by a factor
$\rho_{\mathrm{wlc}}\in [0,1]$: $z = \rho_{\mathrm{wlc}}\,(\ell -
\ell_\mathrm{p})$.
This factor depends on both the pulling force $F_{\mathrm{ext}}$ and
the bending persistence length $A=K_0/(kT)$ and can either be read off
an experimental hat curve from the value $z(n=0) =
\rho_{\mathrm{wlc}}\,\ell$, or computed from
theoretical formulas~\cite{marko+siggia:1995a,bouchiat+al:1999}.
To account for these thermal effects, we replace
Eq.~(\ref{eqn:VerticalExtension}) with:
\begin{equation}
	z =
\rho_{\mathrm{wlc}}\left(1+\frac{2\,K_0}{K_3}\frac{\sin^2\alpha}{\cos2\alpha}\right)\ell
	+\chi \, \rho_{\mathrm{wlc}}\, \frac{4\pi\,R}{\sin2\alpha} \,n~.
	\label{eqn:VerticalExtensionWLC}
\end{equation}

One of the main features of the experimental hat curves is the linear
decrease of the vertical extension with the number of turns.  We
define the slope $q$ in the linear part of the hat curve as:
\begin{equation}
	q \DEF \left| \frac{\d z}{\d n} \right|=\rho_{\mathrm{wlc}}
	\frac{4\pi\,R}{\sin2\alpha}~.
	\label{eqn:slope hat curve}
\end{equation}
Given experimental values of $F_{\mathrm{ext}}$ and $q$,
Eqs.~(\ref{eqn:VarPullingForce}) and~(\ref{eqn:slope hat curve}) can
be solved for $R$ and $\alpha$.  Since $q$ (and $F_{\mathrm{ext}}$)
are constant along the linear part of a hat curve, the values of $R$
and $\alpha$ thus determined will be constant as well.  As a result,
the twisting moment in the molecule, given by
Eq.~(\ref{eqn:VarTorsionalMoment}), is constant, for a given
experiment, along the linear region of the hat curve, a property that
has been previously reported in the
literature~\cite{bouchiat_2000,marko:2007} and which is a clear
outcome of the present model.
An interpretation of the fact that $R$ and $\alpha$ are constant in
the linear region of the hat curve is that each additional turn of
the bead is used to convert a small piece of tail into plectonemes.

\subsection*{Twisting moment}
%
%
%
%
The twisting moment in the molecule, which is uniform and equal to
$M_{\mathrm{ext}}$ at equilibrium, cannot be measured in magnetic
tweezers experiments.
However, it has been shown that enzyme activity such as RNA polymerase
depends on the value of the twisting moment in
DNA~\cite{revyakin+al:2004}.
The value of $M_{\mathrm{ext}}$ can be determined from
Eq.~(\ref{eqn:VarTorsionalMoment}) once $R$ and $\alpha$ are known, as
explained above.
Here, we give a formula for $M_{\mathrm{ext}}$ directly as function of
the experimental slope $q$ and the external force $F_{\mathrm{ext}}$.
Indeed, using Eq.~(\ref{eqn:slope hat curve}) to eliminate $R$ in
Eqs.~(\ref{eqn:VarTorsionalMoment}) and~(\ref{eqn:VarPullingForce}),
one obtains $M_{\mathrm{ext}}(q,\alpha)$ and
$F_{\mathrm{ext}}(q,\alpha)$ as functions of $q$ and $\alpha$.  It is
then possible to eliminate $\alpha$, which yields:
\begin{equation}
	M_{\mathrm{ext}} =
	m+\left( m^2 + 2 K_0 \,F_{\mathrm{ext}} \right)^{1/2}
	,
	\qquad
	\textrm{where }m=
	\frac{q \,F_{\mathrm{ext}}}{4 \pi\, \rho_{\mathrm{wlc}}} 
	- \frac{3\pi\, \rho_{\mathrm{wlc}} \,K_0}{2q}
	\label{eqn:ForceMoment}
\end{equation}
In the limit of small $\alpha$, one can expand the functions
$M_{\mathrm{ext}}(q,\alpha)$ and $F_{\mathrm{ext}}(q,\alpha)$ prior to
elimination of $\alpha$, and this leads to a simplified formula:
\begin{equation}
	M_{\mathrm{ext}} \simeq \frac{2\,q}{3\,\pi \, \rho_{\mathrm{wlc}}}F_{\mathrm{ext}}
	\textrm{,}
	\label{eqn:ApproxForceMoment}
\end{equation}
where, as explained above, $\rho_{\mathrm{wlc}} = z(n=0)/\ell$.  As
shown in Fig.~\ref{fig:TorsionalMoment}, this approximation is
accurate when used with typical experimental values.
Eq.~(\ref{eqn:ApproxForceMoment}) provides a simple and direct mean of
evaluating the twisting moment in magnetic tweezers experiments, based
on the slope of the linear part of the hat curve only.
Note that it should not be inferred from
Eq.~(\ref{eqn:ApproxForceMoment}) that $M_{\mathrm{ext}}$ depends
linearly on $F_{\mathrm{ext}}$, as the slope $q$ is itself a function
of $F_{\mathrm{ext}}$.
\subsection*{Superhelical angle limit}

It is known that the topology of contact between two impenetrable
helical tubes winding along a common axis changes when $\alpha$
becomes larger than $\pi/4$~\cite{neukirch+heijden:2002}.  The
possibility of such a change of topology is not considered in our
model (being specific to hard-core repulsion between tubes, it is not
relevant to DNA molecules undergoing long-range electrostatic
repulsion anyway).  Nevertheless, the equilibrium solutions found here
are all such that $\alpha < \pi/4$.  This upper bound has a mechanical
origin, and not a geometrical one: the expressions for
$F_{\mathrm{ext}}$ in Eq.~(\ref{eqn:VarPullingForce}) and for
$M_{\mathrm{ext}}$ in Eq.~(\ref{eqn:VarTorsionalMoment}) both diverge
at $\alpha=\pi/4$ and plectonemic solutions with a superhelical angle
larger than $\pi/4$ are unstable.

%
%

%
%
%
\subsection*{Application to experiments}
\begin{figure}
    \begin{center}
	\begin{psfrags}
	    \psfrag{n}{$n$} \psfrag{y}{$z\;[\mu\mathrm{m}]$}
	    \includegraphics[width=9.cm]{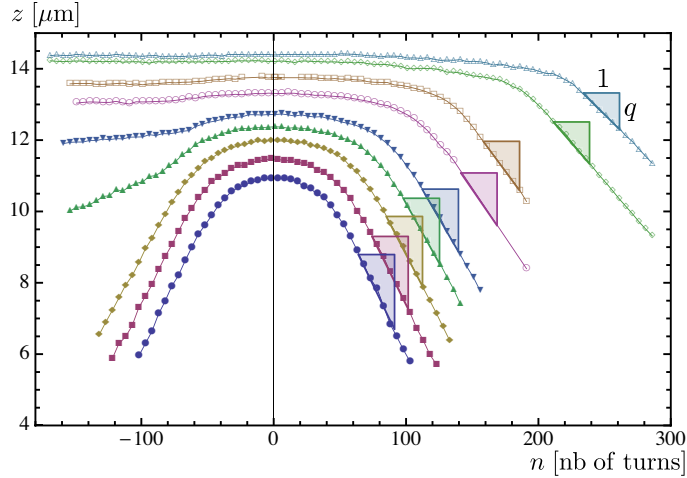}
	\end{psfrags}
	\caption{Experimental hat curves showing the vertical
	extension of a lambda phage DNA 48kbp molecule as a function
	of the number of turns imposed on the magnetic bead (salt
	concentration $10\,\mathrm{mM}$, temperature $298\,K$).
	Experimentally measured persistence length of the molecule is
	$A=51.35\,\mathrm{nm}$.  Each curve corresponds to a fixed
	pulling force $F_{\mathrm{ext}}$: $0.25$, $0.33$, $0.44$,
	$0.57$, $0.74$, $1.10$, $1.31$, $2.20$, $2.95\,\mathrm{pN}$.
	Triangles represent the fit for the slope $q$ of the linear
	region.  Data kindly provided by V. Croquette (LPS -- ENS,
	Paris).}
		\label{fig:HatCurves}
	\end{center}
\end{figure}

%
%
%
%
The model is used to extract mechanical and geometrical parameters
from experimental data.  To allow comparison with previous work, we
use the same data as in~\cite{neukirch_2004}.  These data are shown in
Fig.~\ref{fig:HatCurves}; they were obtained on a 48kbp lambda phage
DNA molecule in a 10mM phosphate buffer.

\begin{figure}
	\begin{center}
		\begin{psfrags}
		\psfrag{r}{$R\;[\mathrm{nm}]$}
		\psfrag{a}{$\alpha\;[\mathrm{rad}]$}
		\psfrag{f}{$F_{\mathrm{ext}}\;[\mathrm{pN}]$}
		\includegraphics[width=9.cm]{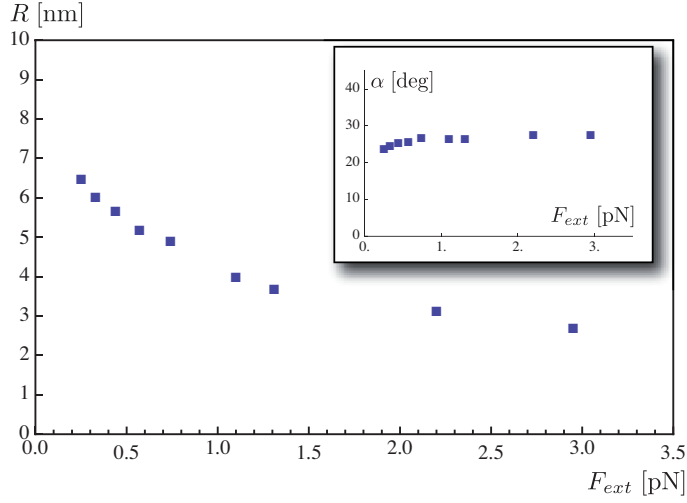}
		\end{psfrags}
		\caption{Reconstructed values of the plectonemic
		radius $R$ as a function of the pulling force, from
		the data in Fig.~\ref{fig:HatCurves} by solving
		Eqs.~(\ref{eqn:VarPullingForce}) and~(\ref{eqn:slope
		hat curve}).  The angle $\alpha$ is shown in the
		inset.}
		\label{fig:PlectonemicVariables}
	\end{center}
\end{figure}

For each curve in Fig.~\ref{fig:HatCurves}, corresponding to a given
value of the external force $F_{\mathrm{ext}}$, we extract the slope
$q$ by fitting the linear region.  The superhelical variables $R$ and
$\alpha$ are found by solving Eqs.~(\ref{eqn:VarPullingForce})
and~(\ref{eqn:slope hat curve}),
and are plotted in Fig.~\ref{fig:PlectonemicVariables} as a function
of $F_{\mathrm{ext}}$.  The reconstructed values of $R$ are in the
nanometric range; they decrease with the pulling force, from
approximately 6 to 2 times the DNA crystallographic radius in this
particular experiment.
At large values of the force, $R$ is close to (and actually smaller
than) the Debye length, $3.07\;\mathrm{nm}$ in $10\;\mathrm{mM}$ salt,
and the Manning condensation radius, $3.18\;\mathrm{nm}$ in
$10\;\mathrm{mM}$ salt~\cite{shaughnessy_2005}.
We note that the values of $R$ found here in the presence of a pulling
force are smaller than (and in the same range as) in
Ref.~\cite{rybenkov+al:1993} where no force is applied, which is
consistent.

\begin{figure}
	\begin{center}
		\begin{psfrags}
		\psfrag{y}{$M_{\mathrm{ext}}\;[\mathrm{pN.nm}]$}
		\psfrag{f}{$F_{\mathrm{ext}}\;[\mathrm{pN}]$}
		\includegraphics[width=9.cm]{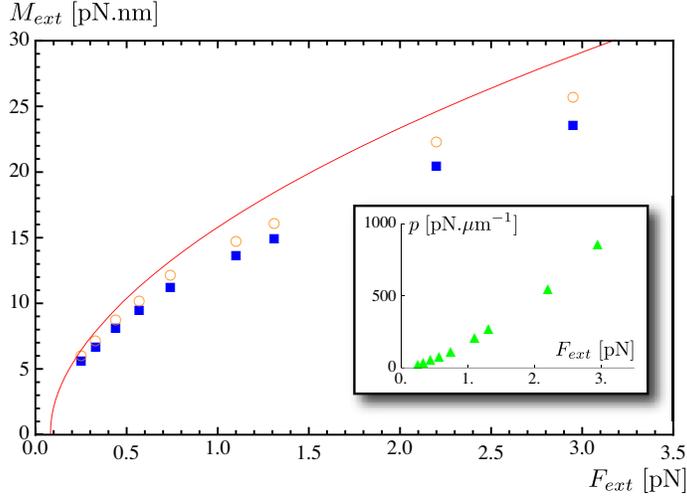}
		\end{psfrags}
		\caption{Reconstructed values for the twisting moment
		in the molecule based on the data shown in
		Fig.~\ref{fig:HatCurves}, using the exact formula in
		Eq.~(\ref{eqn:VarTorsionalMoment}) (solid squares),
		and the small angles approximation in
		Eq.~(\ref{eqn:ApproxForceMoment}) (open circles).
		Comparison with the prediction of the composite model
		in Ref.~\cite{marko:2007} (curve).  Contact pressure
		is shown in inset.}
		\label{fig:TorsionalMoment}
	\end{center}
\end{figure}

The reconstructed values of the twisting moment $M_{\mathrm{ext}}$ and
of the contact pressure $p$ are given in
Fig.~\ref{fig:TorsionalMoment}, based on the same experimental data.
The values of $M_{\mathrm{ext}}$ are determined both by
Eq.~(\ref{eqn:VarTorsionalMoment}) using the previously computed
values of $R$ and $\alpha$, and by the approximate
formula~(\ref{eqn:ApproxForceMoment}) directly.  A good agreement is
obtained, which validates the proposed approximation.  The values of
$M_{\mathrm{ext}}$ are also compared to those predicted by a composite
analytical model, see Eq.~(17) in Ref~\cite{marko:2007} (this model
uses effective parameters determined from Monte-Carlo
simulations~\cite{vologodskii+marko:1997}).
%
%
%
\section{Conclusion}
%
%
%
%
%
We have shown that, under the approximation that thermal fluctuations
are neglected in the plectonemes, one can calculate analytically the
response of twisted DNA: supercoils are described by a mechanically
exact and self-contained model.
Self-contact in the plectonemic region is treated with a hard-core
potential; an expression for the contact pressure between the two
dsDNA is derived.
The hard-core radius is an effective parameter determined, for a given
value of the applied force, from the slope of the linear region of the
experimental curve.
A formula for the twisting moment is proposed, as a function of the
slope of the linear region of the experimental hat curve only.
We apply this analysis to experimental data from which we extract the
mechanical quantities: superhelical radius and angle, contact pressure
and twisting moment.  We compared these values with predictions from
previous analyses, when available, and found that they are consistent.
In future work, we shall extend the present model to deal with
long-range interaction potentials, predict the superhelical radius,
and utilize magnetic tweezers experiments to probe DNA-DNA
electrostatic interaction.
The present paper is a first step towards a mechanically accurate
description of bare dsDNA subjected to tensile and torsional loads, a
problem relevant to the architecture of DNA in the cell nucleus where
proteins come into play.

We thank V. Croquette for allowing us to use his unpublished
experimental data.

\bibliography{dna_pre}

\end{document}